\documentclass[letterpaper,twocolumn,prl,aps,superscriptaddress,amsmath,amssymb,floatfix]{revtex4-1}
\usepackage{mathptmx}
\usepackage[latin9]{inputenc}
\usepackage{color}
\usepackage{verbatim}
\usepackage{float}
\usepackage{amsmath}
\usepackage{amssymb}
\usepackage{graphicx}
\usepackage{esint}
\usepackage[unicode=true,
 bookmarks=true,bookmarksnumbered=false,bookmarksopen=false,
 breaklinks=false,pdfborder={0 0 1},backref=false,colorlinks=true]
 {hyperref}
\hypersetup{
 linkcolor=magenta,urlcolor=blue,citecolor=blue,pdfstartview={FitH},hyperfootnotes=false}

\makeatletter

\usepackage{textcomp}
\usepackage{epstopdf}

\pdfpageheight\paperheight
\pdfpagewidth\paperwidth

\@ifundefined{textcolor}{}{%
 \definecolor{BLACK}{gray}{0}
 \definecolor{WHITE}{gray}{1}
 \definecolor{RED}{rgb}{1,0,0}
 \definecolor{GREEN}{rgb}{0,1,0}
 \definecolor{BLUE}{rgb}{0,0,1}
 \definecolor{CYAN}{cmyk}{1,0,0,0}
 \definecolor{MAGENTA}{cmyk}{0,1,0,0}
 \definecolor{YELLOW}{cmyk}{0,0,1,0}
}

\usepackage{xcolor}
\usepackage{soul}
\definecolor{blue}{rgb}{0,0,1}
\definecolor{red}{rgb}{1,0,0}
\definecolor{green}{rgb}{0,1,0}

\usepackage{soul}

\makeatother

\begin{document}
\title{Synthetic five-wave mixing in an integrated microcavity for visible-telecom
entanglement generation}
\author{Jia-Qi~Wang}
\thanks{These two authors contributed equally to this work.}
\affiliation{CAS Key Laboratory of Quantum Information, University of Science and Technology of China, Hefei 230026, P. R. China.}
\affiliation{CAS Center For Excellence in Quantum Information and Quantum Physics, University of Science and Technology of China, Hefei, Anhui 230026,
P. R. China.}
\author{Yuan-Hao~Yang}
\thanks{These two authors contributed equally to this work.}
\affiliation{CAS Key Laboratory of Quantum Information, University of Science and Technology of China, Hefei 230026, P. R. China.}
\affiliation{CAS Center For Excellence in Quantum Information and Quantum Physics,
University of Science and Technology of China, Hefei, Anhui 230026,
P. R. China.}
\author{Ming~Li}
\email{lmwin@ustc.edu.cn}
\affiliation{CAS Key Laboratory of Quantum Information, University of Science and Technology of China, Hefei 230026, P. R. China.}
\affiliation{CAS Center For Excellence in Quantum Information and Quantum Physics,
University of Science and Technology of China, Hefei, Anhui 230026,
P. R. China.}
\author{Hai-Qi~Zhou}
\affiliation{CAS Key Laboratory of Quantum Information, University of Science and Technology of China, Hefei 230026, P. R. China.}
\affiliation{CAS Center For Excellence in Quantum Information and Quantum Physics,
University of Science and Technology of China, Hefei, Anhui 230026,
P. R. China.}

\author{Xin-Biao~Xu}
\affiliation{CAS Key Laboratory of Quantum Information, University of Science and Technology of China, Hefei 230026, P. R. China.}
\affiliation{CAS Center For Excellence in Quantum Information and Quantum Physics,
University of Science and Technology of China, Hefei, Anhui 230026,
P. R. China.}
\author{Ji-Zhe~Zhang}
\affiliation{CAS Key Laboratory of Quantum Information, University of Science and Technology of China, Hefei 230026, P. R. China.}
\affiliation{CAS Center For Excellence in Quantum Information and Quantum Physics,
University of Science and Technology of China, Hefei, Anhui 230026,
P. R. China.}
\author{Chun-Hua~Dong}
\affiliation{CAS Key Laboratory of Quantum Information, University of Science and Technology of China, Hefei 230026, P. R. China.}
\affiliation{CAS Center For Excellence in Quantum Information and Quantum Physics,
University of Science and Technology of China, Hefei, Anhui 230026,
P. R. China.}

\author{Guang-Can~Guo}
\affiliation{CAS Key Laboratory of Quantum Information, University of Science and Technology of China, Hefei 230026, P. R. China.}
\affiliation{CAS Center For Excellence in Quantum Information and Quantum Physics,
University of Science and Technology of China, Hefei, Anhui 230026,
P. R. China.}

\author{Chang-Ling~Zou}
\email{clzou321@ustc.edu.cn}
\affiliation{CAS Key Laboratory of Quantum Information, University of Science and Technology of China, Hefei 230026, P. R. China.}
\affiliation{CAS Center For Excellence in Quantum Information and Quantum Physics,
University of Science and Technology of China, Hefei, Anhui 230026,
P. R. China.}
\date{\today}
\begin{abstract}
Nonlinear optics processes lie at the heart of photonics and quantum optics for their indispensable role in light sources and information processing. During the past decades, the three- and four-wave mixing ($\chi^{(2)}$ and $\chi^{(3)}$) effects have been extensively studied, especially in the micro-/nano-structures by which the photon-photon interaction strength is greatly enhanced. So far, the high-order nonlinearity beyond the $\chi^{(3)}$ has rarely been studied in
dielectric materials due to their weak intrinsic nonlinear susceptibility, even in high-quality microcavities. Here, an effective five-wave mixing process ($\chi^{(4)}$) is synthesized for the first time, by incorporating $\chi^{(2)}$ and $\chi^{(3)}$ processes in a single microcavity. The coherence of the synthetic $\chi^{(4)}$ is verified by generating time-energy entangled visible-telecom photon-pairs, which requires only one drive laser at the telecom waveband. The photon pair generation rate from the synthetic process shows an enhancement factor over $500$ times upon intrinsic five-wave mixing. Our work demonstrates a universal approach of nonlinear synthesis via photonic structure engineering at the mesoscopic scale rather than material engineering, and thus opens a new avenue for realizing high-order optical nonlinearities and exploring novel functional photonic devices.

\end{abstract}
\maketitle
\noindent \textbf{\large{}Introduction}{\large\par}

\noindent Since the invention of lasers, a wide range of nonlinear optics processes have been experimentally observed in dielectric materials and have deepened our understanding of the physics of light-matter interactions~\citep{Boyd2003,Agrawal2019,ArunKumar2013, Fejer1994, nonlinearrmp}. Nonlinear optics effects not only provide a unique testbed for studying nonlinear physics with flexible parameters
over many orders, but also allow various applications, including frequency conversion for optical detection and imaging~\citep{Guo2020}, comb laser-based precision spectroscopy~\citep{combreview}, material characterization and bio-chemical sensing ~\citep{Tran2017}. For emerging quantum information science, optical nonlinearity is the key resource for nontrivial tasks enabled by quantum mechanics, ranging from communication, sensing to computation. Coherent multi-wave mixing allows the generation of entangled photon pairs~\citep{photonpair} and also quantum frequency conversion~\citep{Kumar1990} to connect distinct quantum systems. In particular, high-order nonlinearities are desired for the generation of exotic many-photon entangled states or the realization of controllable photon-photon quantum gates~\citep{Kok2010,Langford2011}, which are crucial for an extensible quantum system. Nonetheless, most studies are limited to the low-order nonlinearity of dielectric materials, i.e. the $\chi^{(2)}$ and $\chi^{(3)}$ processes, because the nonlinear susceptibility decays exponentially with the order. To date, the fundamental physics of high-order optical nonlinearity and the associated applications are rarely investigated in experiments using solid-state materials~\citep{Ghimire2011,Cox2017}.

Recently, nonlinear photonic devices on integrated photonic chips~\citep{Strekalov2016,Breunig2016,Elshaari2020} have attracted tremendous research interest due to their advantages in compactness, stability, and low-power-consumption. Compared with conventional bulky nonlinear crystals and nonlinear fibers, integrated microcavity
significantly enhances light-matter interaction due to the strongly confined mode volume $V_{\mathrm{m}}$ as well as the high-quality factor~\citep{Xiao2020}. In a microcavity, the nonlinear coupling rate of an $(n+1)$-wave mixing process scales as $g_{n}\propto\chi^{(n)}/V_{\mathrm{m}}^{\left(n-1\right)/2}$, with $\chi^{(n)}\sim\mathcal{O}\left(10^{-10(n-1)}\,\mathrm{(V/m)}^{n-1}\right)$ being the nonlinear susceptibility~\citep{Boyd2003}. Currently, nonlinear enhancement has reduced the optical parametric oscillation threshold dramatically to micro-Watts based on $\chi^{(2)}$ and $\chi^{(3)}$ processes~\citep{Lu2021,Marty2021}. However, the enhancement of $g_{n}$ provided by the scaling factor $(V_{\mathrm{m}})^{(n-1)/2}$ is usually on the order of $10^{5(n-1)}$ for widely studied integrated microcavities, which cannot compensate the $10$ order decrease of high order $\chi^{(n)}$ against $n$. It is still challenging to directly realize the multi-wave mixing involving five or more photons based on the intrinsic nonlinear susceptibility of common materials, even in microcavities.

\noindent
\begin{figure}
\begin{centering}
\includegraphics[width=1\columnwidth]{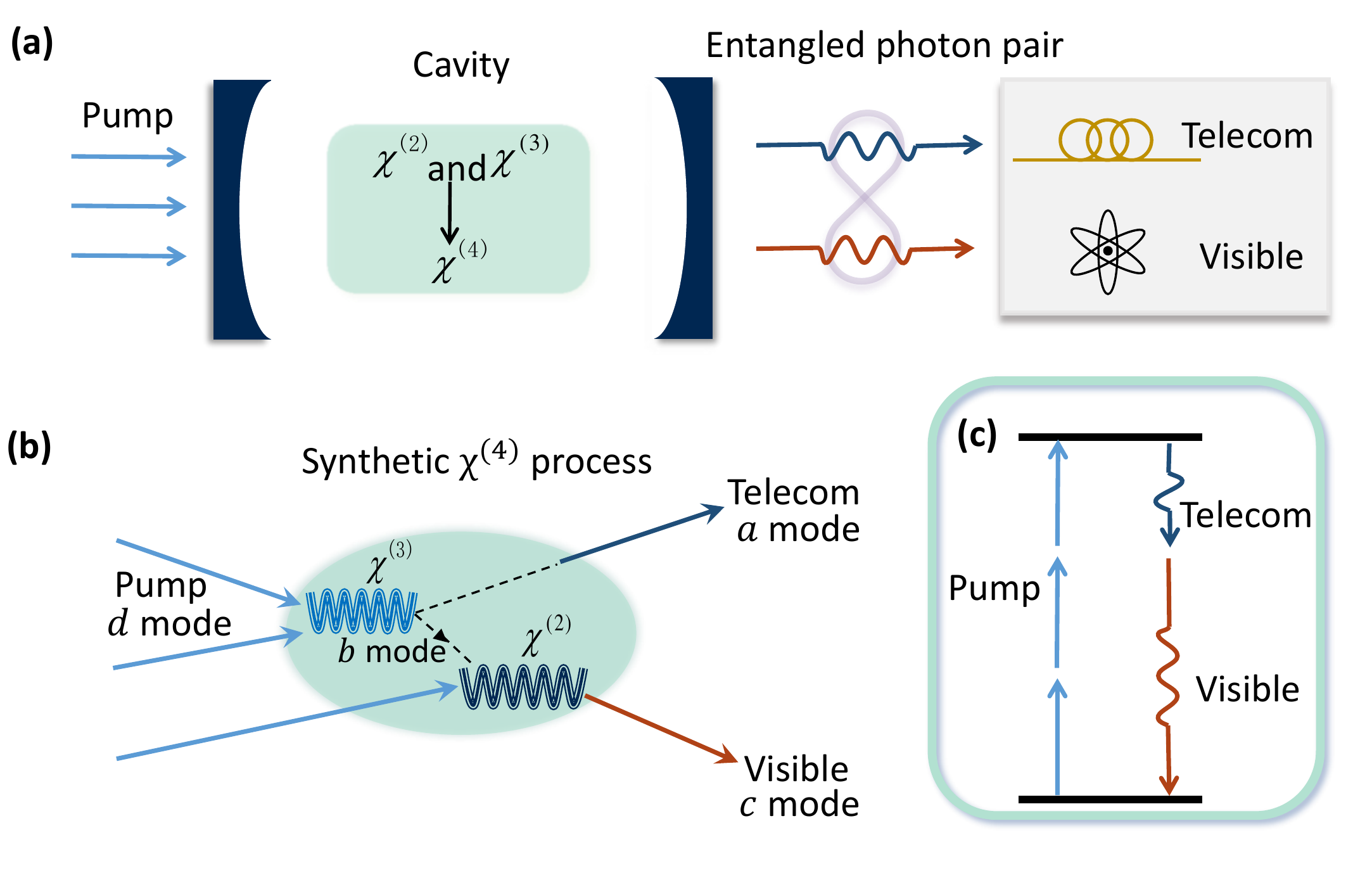}
\par\end{centering}
\caption{\textbf{Schematic illustration of synthetic five-wave mixing.} \textbf{a}, Synthetic $\chi^{(4)}$ process constructed by cavity-enhanced $\chi^{(2)}$ and $\chi^{(3)}$ nonlinear processes for visible-telecom entangled photon-pair generation. \textbf{b},\textbf{c}, The scattering map and the equivalent energy level diagram of the synthetic $\chi^{(4)}$ process.}
\label{Fig1}
\end{figure}

In this work, we demonstrate a novel approach to synthesize $\chi^{(4)}$ process by intrinsic low-order nonlinear processes in a microcavity. The synthetic $\chi^{(4)}$ nonlinearity is demonstrated by generating correlated photon pairs, which shows a rate over $500$ times higher than that due to the intrinsic $\chi^{(4)}$ susceptibility of the material. Its coherent property is verified by measuring the two-photon quantum interference in the time domain, which manifests an entangled photon-pair source between the visible and telecom bands. Compared with schemes based on traditional low-order nonlinear processes that use ultraviolet or specially-designed laser wavelength, our synthetic five-wave mixing approach shows high flexibility in choosing the wavelengths of photons, thus reducing the difficulties in dispersion engineering. Our scheme points to a universal route to synthesize high-order nonlinear processes based on low-order nonlinear processes in a single microcavity and would stimulate future experimental investigations on even higher-order nonlinear processes and the preparation of exotic photonic quantum states.


\smallskip{}

\noindent \textbf{\large{}Results}{\large\par}

\noindent \textbf{Synthetic five-wave mixing}

\noindent The principle of synthetic optical nonlinearity is illustrated
in Fig.$\:$\ref{Fig1}. In an optical cavity filled with non-centrosymmetric
materials {[}Fig.~\ref{Fig1}(a){]}, the low-order nonlinear optical
processes, i.e. three-wave mixing (3WM) and four-wave mixing (4WM)
due to the material's intrinsic $\chi^{(2)}$ and $\chi^{(3)}$ nonlinearities, respectively, could both be enhanced by the resonances. The two separate processes are independent of each other except sharing
a common optical mode. For instance, as shown by the scattering map
in Fig.$\:$\ref{Fig1}(b), a photon generated by the 4WM could be
a seed for the 3WM, and eventually an effective five-wave mixing (5WM)
is synthesized by incorperating 4WM and 3WM. Denoting the shared mode for
4WM and 3WM as $b$, the four-photon and three-photon interactions
could be described by the Hamiltonian as $g_{3}\left(abd^{\dagger2}+a^{\dagger}b^{\dagger}d^{2}\right)$
and $g_{2}\left(b^{\dagger}d^{\dagger}c+bdc^{\dagger}\right)$, respectively,
where $a,c,d$ represent the other involved photonic modes, and $g_{2(3)}$
represents the photon-photon interaction strength due to second (third)-order
nonlinearity. When accessing the system only through modes $a$, $c$
and $d$, and by treating the intermediate photon in $b$ as virtual
excitation, we could obtain the synthetic 5WM as
\begin{eqnarray}
H & = & g_{\mathrm{4,eff}}\left(d^{\dagger3}ac+a^{\dagger}c^{\dagger}d^{3}\right),\label{eq:fiveHam}
\end{eqnarray}
where $g_{4,\mathrm{eff}}=g_{2}g_{3}/\Lambda_{b}$ is the effective five-photon interaction strength and $\Lambda_{b}$ is the equivalent detuning of the intermediate mode $b$ {[}see Supplementary Information for more details{]}.
Note that $g_{4,\mathrm{eff}}$ could be a complex number, which indicates
a non-Hermitian 5WM process and more interesting physics about this
is left for future studies. The synthetic 5WM can be used to construct the parametric
interaction between telecom wavelength modes ($d$ and $a$)
and visible mode $c$: when driving $d$, a pair of photon could be
generated in $a$ and $c$, as depicted in Fig.~\ref{Fig1}(c). Compared
with the material's intrinsic $\chi^{(4)}$, the synthetic nonlinearity
holds many advantages: First, the synthetic $g_{4,\mathrm{eff}}$
could be much higher than that from intrinsic $\chi^{(4)}$, thus enabling
stronger multi-photon interactions. Second, elementary
(low-order) process could be engineered individually and then combined for synthetic
nonlinearity, thus the complicated dispersion engineering for high-order
modes with poor modal overlaps is avoided for practical applications.
It also provides an universal approach for constructing higher-order
processes. For the example above, if we choose $d$ as the virtual photons instead, six-wave mixing $g_{5,\mathrm{eff}}\left(ab^{3}c^{\dagger2}+h.c.\right)$ could be realized ($g_{5,\mathrm{eff}}\propto g_{3}g_{2}^{2}$) by combining one $\chi^{(3)}$ and two $\chi^{(2)}$ processes (see Supplementary Information).

\begin{figure*}[t]
\begin{centering}
\textcolor{red}{\includegraphics[width=1\textwidth]{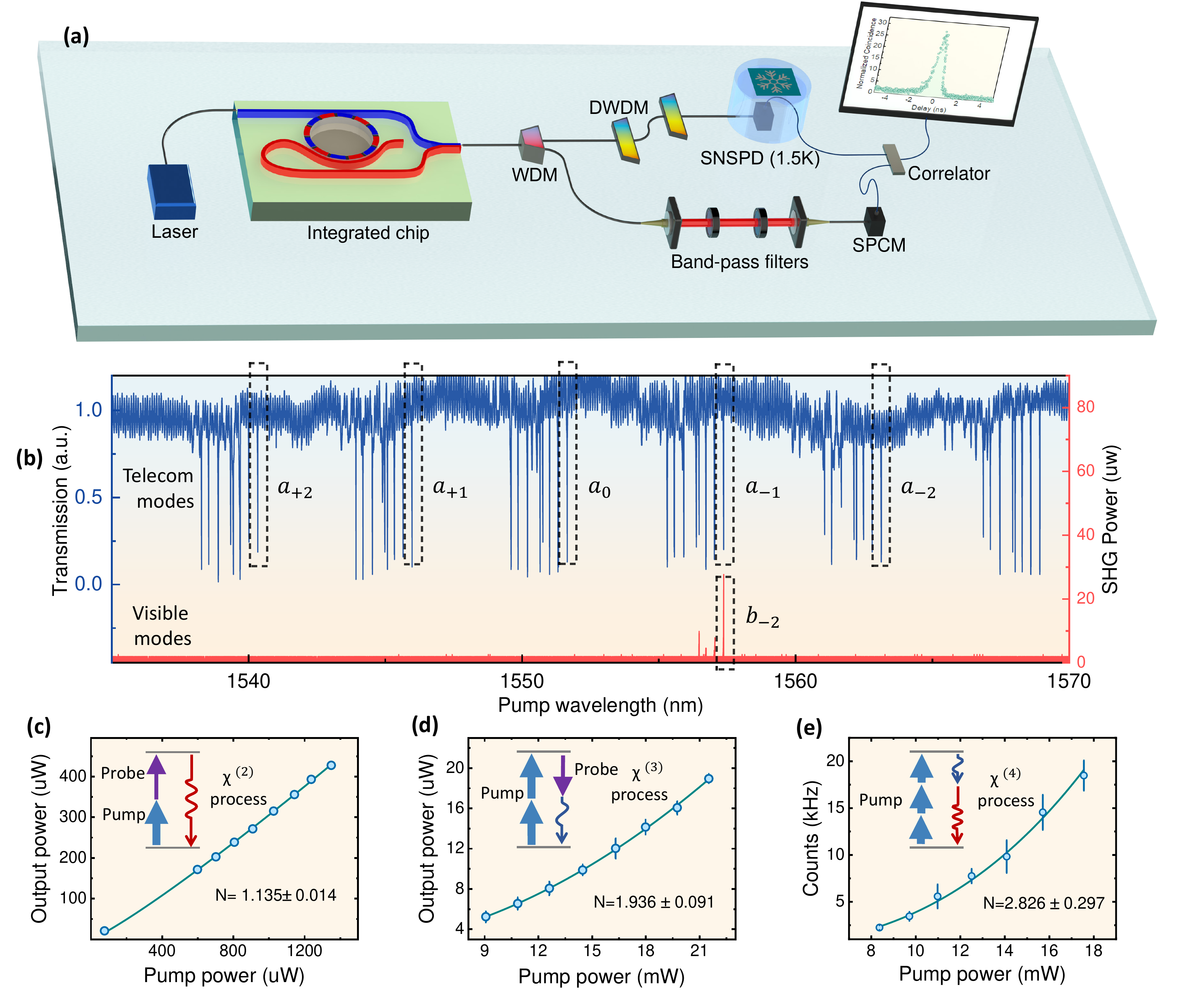}}
\par\end{centering}
\caption{\textbf{Characterization of the synthetic nonlinear process.} \textbf{a}, The experimental setup. Telecom pump and probe lasers are coupled into the chip through a fiber lens and the output signals are collected by another fiber lens. The visible and telecom signal photons are separated by a wavelength division multiplexer (WDM), and then filtrated from background via a series of band-pass filters and dense wavelength division multiplexing (DWDM), respectively. The coincidence of generated photon pairs is characterized by single photon detectors (SPCM and SNSPD) and a correlator. \textbf{b}, Transmission spectrum of telecom modes and the corresponding second-harmonic generation. The dips in dashed frames correspond to the modes characterized for  synthetic five-wave mixing. The on-chip pump power is $1.8\,{\rm mW}$. \textbf{c-e}, The input-output relation of $\chi^{(2)}$, $\chi^{(3)}$ and $\chi^{(4)}$ processes. The on-chip probe power in \textbf{c} is $22\,{\rm mW}$ and in \textbf{d} is $1.1\,{\rm mW}$. Error bars denote standard deviations.}
\label{Fig2}
\end{figure*}

\smallskip{}

\noindent \textbf{Experimental characterization}

\noindent The 5WM process is experimentally synthesized in a chip-integrated aluminum
nitride microring, which provides excellent $\chi^{(2)}$
and $\chi^{(3)}$ properties and has been extensively studied in comb
generation and high-efficient second harmonic generation (SHG)~\citep{Guo2016,Bruch2021}.
Here, the effective $\chi^{(4)}$ is constructed between the fundamental
$\mathrm{TM}_{00}$ modes at telecom wavelength($\sim1550\,\mathrm{nm}$) and $\mathrm{TM}_{20}$ modes at visible wavelength ($\sim775\,\mathrm{nm}$). For the telecom modes only in a relatively narrow frequency range ($\sim1550\pm15\,\mathrm{nm}$),
the dispersion is negligible and all modes could participate in
the 4WM efficiently. The microring was subsequently designed with
an appropriate width to realize the phase-matching of 3WM between
the $\mathrm{TM_{00}}$ telecom drive mode and $\mathrm{TM_{20}}$
visible mode. Efficient sum-frequency generation (SFG) is realized by finely tuning the chip temperature near the SFG phase-matching point. As the experimental setup shows in Fig.$\,$\ref{Fig2}(a), telecom drive laser and input signals are injected into the device from one side of the chip through a fiber lens. The output signals are collected by another fiber lens on the other side of the chip, and are sequentially separated into different paths by a wavelength division multiplexer (WDM). After that, we use single photon counting modules (SPCM) to detect visible output signal, with the potential background noise filtered by a series of band-pass filters, while the telecom output signal transmitted through cascaded dense wavelength division multiplexings (DWDM) is detected by a superconducting nanowire single photon detector (SNSPD).

Figure$\,$\ref{Fig2}(b) shows the transmission spectrum of the telecom
modes, where the modes belonging to the same mode family are marked
by black frames. Along with the scanning of the laser, a strong peak in
the visible mode is observed as the phase-matching condition between modes $a_{-1}$ and $b_{-2}$
is satisfied, which indicates the highly efficient SHG.
Here, $a_{i},b_{i}$ denote the bosonic operators of
the optical modes, with the subscript $i\in\mathbb{Z}$ denoting the
relative angular momentum of the modes.
Due to the small dispersion of telecom modes $\omega_{a,-2}+\omega_{a,0}\approx2\omega_{a,-1}$,
the SHG also implies an efficient non-degenerate 3WM process (i.e. SFG)
$H_{\mathrm{SFG}}=g_{2}(a_{0}a_{-2}b_{-2}^{\dagger}+h.c.)$. Combining the SFG
with the special 4WM process $H_{\mathrm{4WM}}=g_{3}(a_{0}a_{0}a_{+2}^{\dagger}a_{-2}^{\dagger}+h.c.)$,
which shares the same telecom mode $a_{-2}$,
the desired 5WM between $a_{0}$, $a_{+2}$ and $b_{-2}$ is constructed
under fast dynamics of the virtual mode $a_{-2}$. This synthetic process consumes three pump photons ($a_{0}$ mode) to produce a telecom-visible photon pair at $a_{+2}$ and $b_{-2}$, respectively.
\begin{figure*}[t]
\begin{centering}
\includegraphics[width=0.9\textwidth]{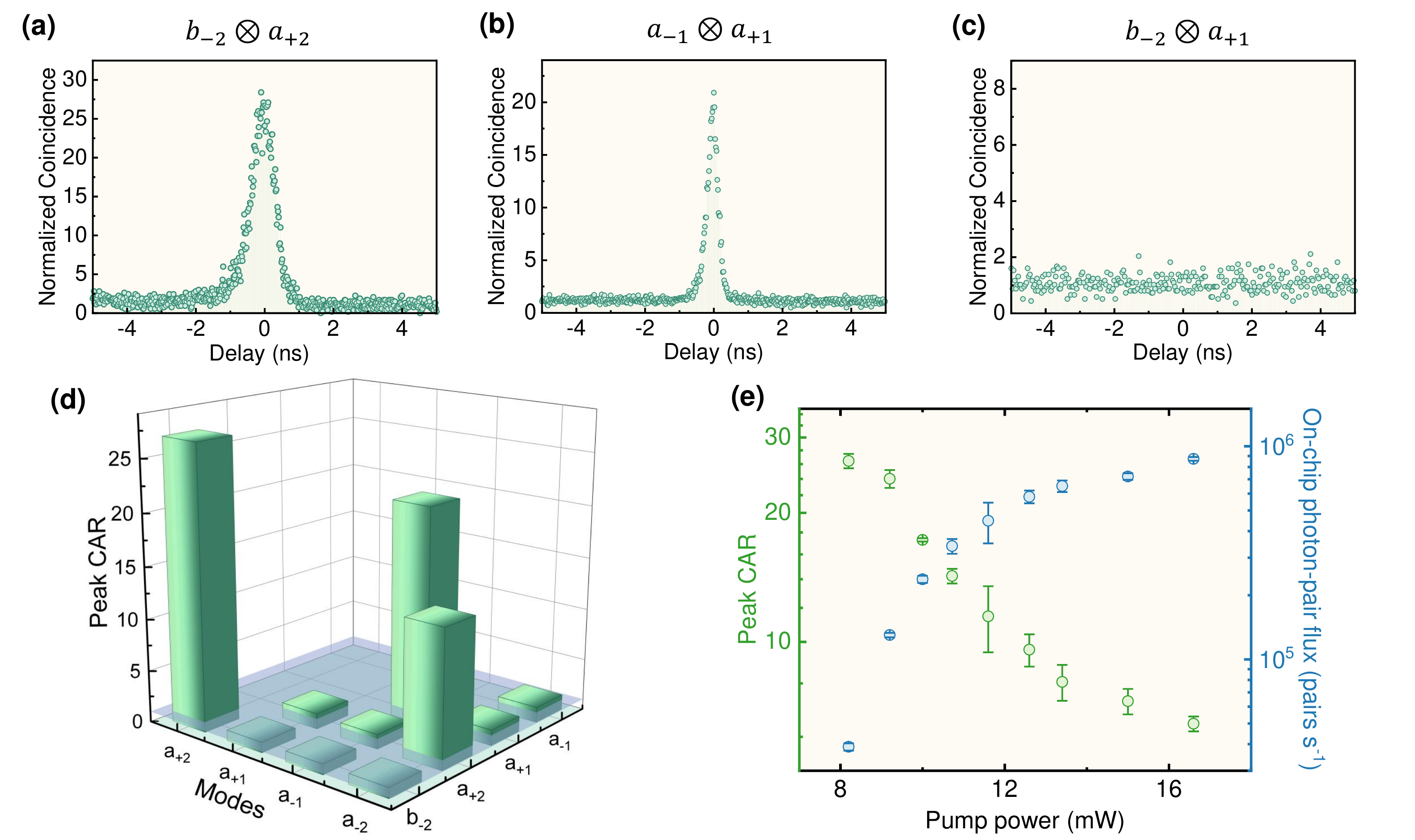}
\par\end{centering}
\caption{\textbf{Photon pair coincidence.} \textbf{a-c}, Normalized coincidence spectra for different mode pairs. The on-chip pump power is $8.2\,{\rm mW}$. \textbf{d}, The peak coincidence to accidental rate (CAR) matrix of different mode pairs. The value of the blue reference flat denotes 1. \textbf{e}, Pump power dependence of the peak CAR for visible--telecom photons ($b_{-2}\bigotimes a_{+2}$) and on-chip photon pair flux. Error bars denote standard deviations.}
\label{Fig3}
\end{figure*}

Inside the AlN microcavity, all these nonlinear optical processes
can be greatly enhanced due to the strong mode confinement and high
quality factor. To ensure
an efficient 5WM process, the elementary nonlinear processes
are verified by pumping the $a_{0}$ and probing the virtual mode
$a_{-2}$. According to $H_{\mathrm{3WM}}$ and $H_{\mathrm{4WM}}$,
for a given probe power, the powers of generated signals from mode
$b_{-2}$ and $a_{+2}$ scale linearly and quadratically with the pump
power for 3WM and 4WM, respectively. Figures$\,$\ref{Fig2}(c)-(d)
show the relation between the output power from the signal mode and the
pump power. By fitting the input-output relationship with $P_{\mathrm{out}}=A\times P_{\mathrm{pump}}^{N}$
with a fixed probe power, we get $N=1.135\,\pm0.014$ for mode $b_{-2}$
and $N=1.936\,\pm0.091$ for mode $a_{+2}$, which proves the efficient
3WM and 4WM associating with the virtual mode. Then, by turning off the
probe laser of the virtual mode, the synthetic 5WM is tested using a
coherent pump on the telecom mode $a_{0}$, which produces an effective
parametric Hamiltonian $g_{\mathrm{eff}}n_{a}^{3/2}\left(b_{-2}a_{+2}+b_{-2}^{\dagger}a_{+2}^{\dagger}\right)$
with an intracavity pump photon number $n_{a}$. Such vacuum-induced
photon-pair generation process results in a single-photon count rate of tens kilo-Herz from the visible mode $b_{-2}$ by the SPCM. The
power-dependent counts are fitted with $N=2.826\,\pm0.297$, agreeing
well with the theoretical prediction of cubic power dependence ($\propto n_{a}^{3}$).

\smallskip{}
\noindent \textbf{Visible-telecom entanglement}

\noindent Although the cubic power dependence of the photon-pair rate demonstrates that five
photons participate simultaneously in this 5WM process, it lacks direct evidence on
the coherence of the synthetic nonlinear process, which is vital for
potential quantum device applications in quantum information processing~\citep{Kwiat1995,Ma2012,Kartik2019},
including the quantum frequency conversion and entangled photon sources.
Therefore, the inherent coherent property of the synthetic $\chi^{(4)}$
process is further investigated by measuring the temporal correlation function and time-energy quantum entanglement between modes at the telecom band and the visible band, under the pump on mode $a_{0}$ by a monochromatic
laser.

Figures.$\,$\ref{Fig3}(a)-(c) show the normalized coincidence spectra
for the visible-telecom correlation ($b_{-2}\bigotimes a_{+2}$) and
($b_{-2}\bigotimes a_{+1}$), as well as telecom-telecom correlation
($a_{-1}\bigotimes a_{+1}$). It is found that only the 4WM and target
5WM produce the correlation, while the un-phase-matched interaction
is excluded. Furthermore, the peak coincidence to accidental rates (CARs) between the involved modes are
summarized in Fig.$\,$\ref{Fig3}(d). The coincidence map shows
that there are correlations between the $a_{+2}$ mode and $a_{-2}$
mode, $a_{-1}$ mode and $a_{+1}$ mode, $a_{+2}$ mode and $b_{-2}$
mode. The former two correlations correspond to the 4WM process and the third correlation correspond to the synthetic 5WM process. These results
unambiguously exclude other potential multi-photon processes and noises. Note that for a Hermitian $\chi^{(4)}$-process, the photon
generation in $a_{-2}$ should be suppressed and thus the coincidence
between $a_{-2}$ and $a_{+2}$ vanishes, which can be achieved in principle
by independently engineering the resonance of mode $a_{-2}$ far-off
the phase-matching. The power dependence of the on-chip pair flux and
peak CAR is shown in Fig.$\,$\ref{Fig3}(e),
with the on-chip photon pair flux derived from the detected photon pair
flux by taking account of the losses of both signal and idler photons. As the pump
power increases, the pair flux increases but the CAR value is limited
by multi-pair generation in our process. While at the low pump region
the decrease of pair flux is accompanied by the suppression of multi-pair
events, so that CAR value is mainly limited by detector dark counts
and imperfect filtering which is caused by the leakage of pump light
at telecom band and pump SHG signal at visible band.

Furthermore, it can be inferred from Eq.$\:$(\ref{eq:fiveHam}) that
a coherent quantum process enables the simultaneous generation of
a pair of photons in target modes $b_{-2}$ and $a_{+2}$, instead of
consequent realization of the 4WM and 3WM in a single microring. Comparing
Fig.~\ref{Fig3}(a) and \ref{Fig3}(b), the temporal correlation
function shows a similar symmetric profile, thus confirming that the
synthetic nonlinearity resembles the intrinsic $\chi^{(3)}$ that
generate photons simultaneously. Due to the higher dissipation rate
of visible mode, the correlation function for $b_{-2}\bigotimes a_{+2}$
shows a spread distribution of the generation time of the photon
pairs, which can be characterized by coincidence measurements of different
time offsets.

\begin{figure}
\begin{centering}
\includegraphics[width=1\columnwidth]{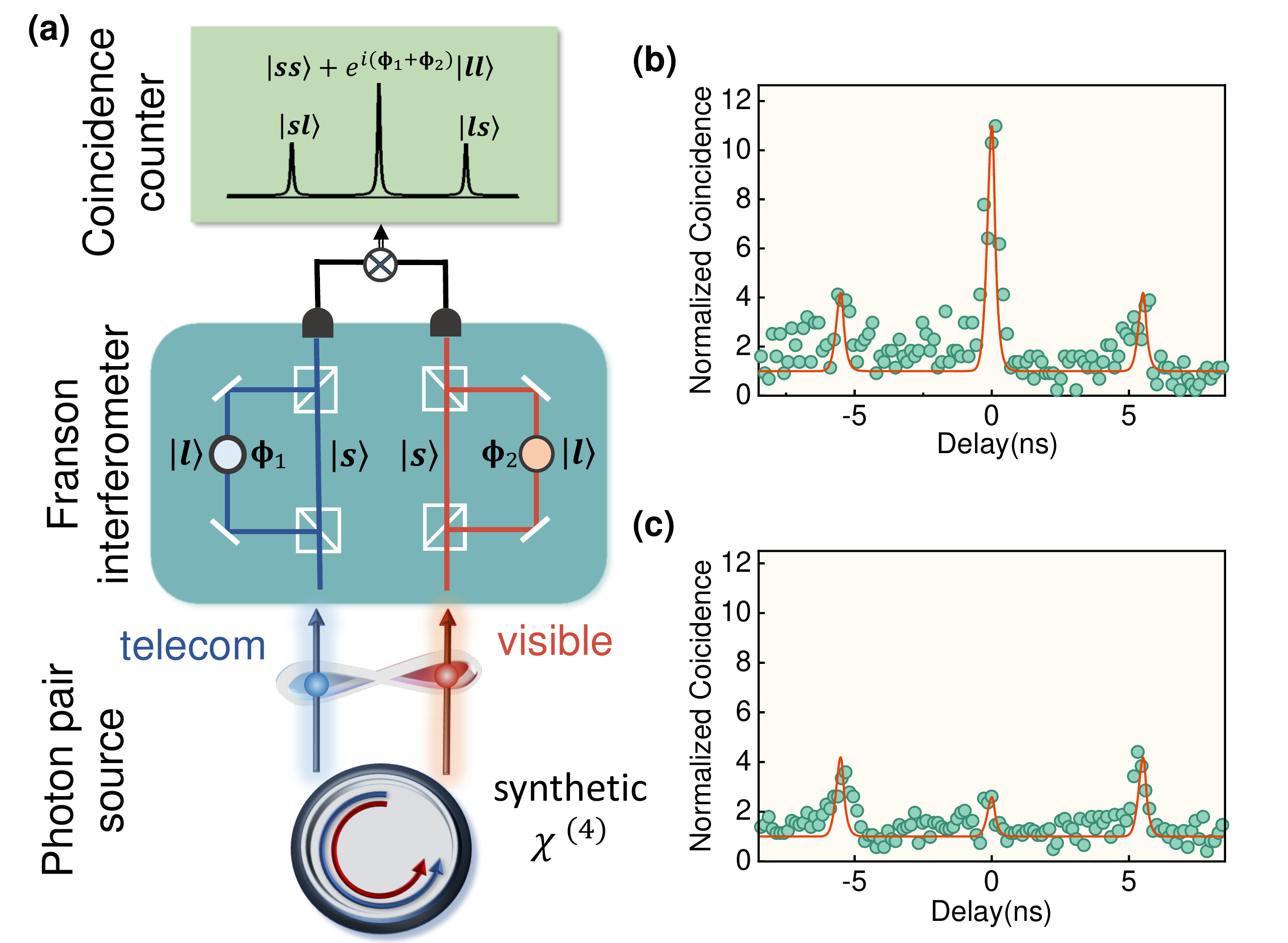}
\par\end{centering}
\caption{\textbf{Visible-telecom entangled photon pair.} \textbf{a}, Characterization of the entanglement via Franson interferometer. There are two-unbalanced Mach--Zehnder interferometers (MZIs), for visible and telecom bands separately. The setup measures the interference between the short--short state ($|ss\rangle$) and long--long state ($|ll\rangle$), with the output coincidence depending on the phases of the MZIs ($\phi_{1}+\phi_{2}$). \textbf{b},\textbf{c}, The interference fringes of the Franson interferometer, showing the constructive and destructive interference, with a  peak CAR visibility of $72.7\%\pm3.3\%$.}
\label{Fig4}
\end{figure}

To verify the quantum coherence property of the synthetic $\chi^{(4)}$
and explore its potential applications, the time-energy entanglement
between the emitted visible-telecom photon pairs is demonstrated.
Through the two unbalanced Mach-Zehnder interferometers (MZIs), the
photon pairs are divided into four paths namely, short--short, long--long,
short--long and long--short for visible--telecom channels, corresponding
to twin-photon amplitude of different times. The time-energy entanglement is characterized via the Franson interferometer~\citep{Franson1989}. As shown at the top of Fig.$\,$\ref{Fig4}(a), both the short-short and long-long twin-photon states contribute to the center peak of the coincidence spectrum, thus interference is expected for coherent parametric interaction. The equal time twin-photon quantum state can
be expressed as: $|\Psi\rangle=|ss\rangle+e^{i(\phi_{1}+\phi_{2})}|ll\rangle$,
where $|ss\rangle$ and $|ll\rangle$ stand for twin photons from
the short--short and long--long arms of the unbalanced MZI. The amplitude
of the coincidence interference peak depends on the phases ($\phi_{1},\phi_{2}$)
of the two MZIs. In our experiment, we tune  the phase $\phi_{1}+\phi_{2}$ and track the coincidence
spectrum. The remarkable change of the center peak in Figs.$\,$\ref{Fig4}(b)-(c)
demonstrates the existence of quantum interference. In particular,
the visibility of the amplitude at the center peak achieves $72.7\%\pm3.3\%$,
which violates the Bell's inequality and manifests a visible-telecom
entangled photon-pair source.

\smallskip{}

\noindent \textbf{\large{}Discussion}{\large\par}

\noindent In summary, we have demonstrated synthetic five-wave mixing in an integrated aluminum nitride microcavity. The coherent and quantum nature of the synthetic nonlinear process is validated by the quantum entanglement between visible and telecom photons, which are generated by a single telecom pump laser. It could be applied as quantum interfaces for future hybrid quantum network based on the Rubidium-atom quantum memories~\citep{Yu2020}. Comparing with the intrinsic $\chi^{(4)}$ nonlinearity of the material, our approach shows an enhancement of photon-pair generation rate by more than 500 times (see Supplementary Information for derivation). In addition, coherent and fast tuning of the synthetic nonlinearity is enabled by controlling the intermediate mode, instead of the material property~\citep{shan2021giant}. Remarkably, this synthetic five-wave mixing also promises efficient on-chip three-photon sources by pumping at visible and telecom wavelengths, as a reversal of the process demonstrated here, and provides the Greenberger-Horne-Zeilinger resource states for photonic fusion-based quantum computing~\citep{Llewellyn2020}. It is anticipated that even higher-order synthetic nonlinearity could be achieved in the lithium niobate and gallium arsenide microcavities~\citep{Kuo2014,ChenLN2019,Lu2020}. Moreover, our approach can be applied to all kinds of nonlinear platforms rather than limited in non-centrosymmetric materials with $\chi^{(2)}$ nonlinearity, and can be extended to hybrid systems, such as acoustic-optics involving photon-phonon interactions. Our demonstration opens the possibility of studying the fundamental physics in nonlinear multi-wave mixing and exploiting new functional quantum photonic devices.

\smallskip{}

\clearpage{}

\noindent \textbf{\large{}Methods}{\large\par}

\noindent \textbf{Experimental device and setup.} The aluminum nitride photonic chip used in our experiment is optimized for high-efficiency second-harmonic generation (SHG). The microring is designed with an appropriate width to realize phase-matching between the $\mathrm{TM_{00}}$
telecom drive mode and $\mathrm{TM_{20}}$ visible mode, and the exact frequency matching between visible and telecom modes is then realized by finely tuning the chip temperature. The radius of our device is approximately 30$\,\mathrm{\mu m}$.  Our device uses a straight bus waveguide to couple telecom light into the microrings and uses a specially-designed wrap-around waveguide for coupling visible light out.

In our experiment, the pump light is provided by amplifying the output of a telecom laser source (Agilent 8164A) with an erbium-doped optical fiber amplifier (EDFA, CONQUER, KG-EDFA-P). The amplified light is transmitted through a dense wavelength division multiplexing (DWDM) to filter the background noise caused by EDFA.  The probe light is New Focus diode laser (TLB-6700) with a tunable laser controller. Our device is placed in an external heater (Covesion, PPLN Ovens-PV10) with a precise temperature controller (Covesion, OC2). For the filter system, we adopt multiple narrow band-pass filters (Semerock, LL01-780) for visible light, which can help us to filter the pump SHG signal and block background noise. In the telecom band, we use commercial 1550~nm band fiber DWDMs, with 100~GHz channel bandwidth.

\vbox{}

\noindent \textbf{Single photon detection.} Two kinds of single photon
detectors were used to detect the generated visible and telecom single-photon-level signals separately. A superconducting nanowire single-photon detector (SNSPD, PHOTEC-1550) is used for telecom outputs with a high detection efficiency over $90\%@1550\,\mathrm{nm}$, a low dark count rate less than $100\,{\rm Hz}$ and a small jitter less than $20\,{\rm ps}$. The single photon counting modules (SPCM, Excelitas SPCM-NIR), with a photon detection efficiency of $70\%@780\,{\rm nm}$ and less than $200\,{\rm Hz}$ dark count rate, is used for output in the visible band. For coincidence measurement between the modes, we use  a high resolution time-to-digital converter (quTAG, standard 4 channels).The peak CAR values in Fig.$\,$\ref{Fig4} are calculated by ${\rm CAR}=(C-A)/A$, where $C$ and $A$ are the overall and accidental coincidence counts obtained from the peak and background of the coincidence counting spectra. The one standard deviation uncertainty in Fig.$\,$\ref{Fig3}(d) is given by multiple measurements.

\vbox{}

\noindent \textbf{Franson interference.} For the Franson interferometer, we use two-unbalanced Mach--Zehnder interferometers (MZIs) for visible and telecom band signals, separately. The visible band unbalanced MZI is made up of spatial optical setup and uses a piezoelectric transducer (PZT) to compensate for the phase drift due to the thermal effect and instability of the MZI via a proportion integration differentiation (PID) controller (SRS-SIM960). A fiber phase shifter and a PID controller are adopted to stabilize and adjust the phase at telecom band.

The interference visibility is calculated by the ratios extracted from the peak of coincidence spectra in Fig.$\,$\ref{Fig4}(b)-(c) by ${\rm Visibility}={\rm \frac{CAR_{max}-CAR_{min}}{CAR_{max}+CAR_{min}}}$, where ${\rm CAR_{max}}$ and $\mathrm{CAR_{min}}$ are the center peak CAR values of the spectra. The one standard deviation uncertainty is given by $\frac{\sigma_{{\rm CAR}}}{{\rm CAR}}\approx\frac{1}{\sqrt{N}}$, $N$ is the total coincidence counts at the peak, and the uncertainty of visibility is derived via the error propagation formula.

The interference visibility is limited by the signal-to-noise ratio of the single-photon outputs, which is mainly attributed to: (i) The low fiber-to-chip coupling efficiency for visible light, which limits the counts of the single photons at the visible wavelengths; (ii) The background noise due to the residue of the pump field and its second-harmonics. It can be improved with high-performance filters or by choosing the signal mode far from the pump in future experiments. (iii) Other parasitical nonlinear effects, such as Raman scattering.

\smallskip{}

\noindent \textbf{\large{}Acknowledgment}{\large\par}

\noindent This work was funded by the National Key Research and Development Program (Grant No.~2017YFA0304504) and the National Natural Science Foundation of China (Grant Nos.~11874342, 11904316, 11922411, and 12104441) and Natural Science Foundation of Anhui Province (Grant No. 2008085QA34 and 2108085MA22). ML and CLZ was also supported by the Fundamental Research Funds for the Central Universities, and the State Key Laboratory of Advanced Optical Communication Systems and Networks. This work was partially carried out at the USTC Center for Micro and Nanoscale Research and Fabrication.

\smallskip{}

\noindent \textbf{\large{}Author contributions}{\large\par}

\noindent C.-L.Z. and M.L. conceived the experiments. J.-Q.W. and Y.-H.Y. built the experimental setup, carried out the measurements, and analyzed the data, with assistance from H.-Q.Z., X.-B.X., J.-Z.Z., and C.-H.D.. M.L. provided theoretical supports. J.-Q.W.,Y.-H.Y., M.L., and C.-L.Z. wrote the manuscript, with input from all other authors. M.L., C.-L.Z., and G.-C.G. supervised the project.


\begin{thebibliography}{32}%
\makeatletter
\providecommand \@ifxundefined [1]{%
 \@ifx{#1\undefined}
}%
\providecommand \@ifnum [1]{%
 \ifnum #1\expandafter \@firstoftwo
 \else \expandafter \@secondoftwo
 \fi
}%
\providecommand \@ifx [1]{%
 \ifx #1\expandafter \@firstoftwo
 \else \expandafter \@secondoftwo
 \fi
}%
\providecommand \natexlab [1]{#1}%
\providecommand \enquote  [1]{``#1''}%
\providecommand \bibnamefont  [1]{#1}%
\providecommand \bibfnamefont [1]{#1}%
\providecommand \citenamefont [1]{#1}%
\providecommand \href@noop [0]{\@secondoftwo}%
\providecommand \href [0]{\begingroup \@sanitize@url \@href}%
\providecommand \@href[1]{\@@startlink{#1}\@@href}%
\providecommand \@@href[1]{\endgroup#1\@@endlink}%
\providecommand \@sanitize@url [0]{\catcode `\\12\catcode `\$12\catcode
  `\&12\catcode `\#12\catcode `\^12\catcode `\_12\catcode `\%12\relax}%
\providecommand \@@startlink[1]{}%
\providecommand \@@endlink[0]{}%
\providecommand \url  [0]{\begingroup\@sanitize@url \@url }%
\providecommand \@url [1]{\endgroup\@href {#1}{\urlprefix }}%
\providecommand \urlprefix  [0]{URL }%
\providecommand \Eprint [0]{\href }%
\providecommand \doibase [0]{http://dx.doi.org/}%
\providecommand \selectlanguage [0]{\@gobble}%
\providecommand \bibinfo  [0]{\@secondoftwo}%
\providecommand \bibfield  [0]{\@secondoftwo}%
\providecommand \translation [1]{[#1]}%
\providecommand \BibitemOpen [0]{}%
\providecommand \bibitemStop [0]{}%
\providecommand \bibitemNoStop [0]{.\EOS\space}%
\providecommand \EOS [0]{\spacefactor3000\relax}%
\providecommand \BibitemShut  [1]{\csname bibitem#1\endcsname}%
\let\auto@bib@innerbib\@empty
\bibitem [{\citenamefont {Boyd}(2003)}]{Boyd2003}%
  \BibitemOpen
  \bibfield  {author} {\bibinfo {author} {\bibfnamefont {R.~W.}\ \bibnamefont
  {Boyd}},\ }\href {\doibase 10.1016/B978-0-12-121682-5.X5000-7} {\emph
  {\bibinfo {title} {Nonlinear Optics}}}\ (\bibinfo  {publisher} {Elsevier},\
  \bibinfo {year} {2003})\BibitemShut {NoStop}%
\bibitem [{\citenamefont {Agrawal}(2019)}]{Agrawal2019}%
  \BibitemOpen
  \bibfield  {author} {\bibinfo {author} {\bibfnamefont {G.~P.}\ \bibnamefont
  {Agrawal}},\ }\href {\doibase 10.1016/C2018-0-01168-8} {\emph {\bibinfo
  {title} {{Nonlinear Fiber Optics}}}}\ (\bibinfo  {publisher} {Elsevier},\
  \bibinfo {year} {2019})\BibitemShut {NoStop}%
\bibitem [{\citenamefont {{Arun Kumar}}(2013)}]{ArunKumar2013}%
  \BibitemOpen
  \bibfield  {author} {\bibinfo {author} {\bibfnamefont {R.}~\bibnamefont
  {{Arun Kumar}}},\ }\bibfield  {title} {\enquote {\bibinfo {title} {{Borate
  Crystals for Nonlinear Optical and Laser Applications: A Review}},}\ }\href
  {\doibase 10.1155/2013/154862} {\bibfield  {journal} {\bibinfo  {journal} {J.
  Chem.}\ }\textbf {\bibinfo {volume} {2013}},\ \bibinfo {pages} {154862}
  (\bibinfo {year} {2013})}\BibitemShut {NoStop}%
\bibitem [{\citenamefont {Fejer}(1994)}]{Fejer1994}%
  \BibitemOpen
  \bibfield  {author} {\bibinfo {author} {\bibfnamefont {M.~M.}\ \bibnamefont
  {Fejer}},\ }\bibfield  {title} {\enquote {\bibinfo {title} {{Nonlinear
  Optical Frequency Conversion}},}\ }\href {\doibase 10.1063/1.881430}
  {\bibfield  {journal} {\bibinfo  {journal} {Phys. Today}\ }\textbf {\bibinfo
  {volume} {47}},\ \bibinfo {pages} {25} (\bibinfo {year} {1994})}\BibitemShut
  {NoStop}%
\bibitem [{\citenamefont {Bloembergen}(1982)}]{nonlinearrmp}%
  \BibitemOpen
  \bibfield  {author} {\bibinfo {author} {\bibfnamefont {N.}~\bibnamefont
  {Bloembergen}},\ }\bibfield  {title} {\enquote {\bibinfo {title} {Nonlinear
  optics and spectroscopy},}\ }\href {\doibase 10.1103/RevModPhys.54.685}
  {\bibfield  {journal} {\bibinfo  {journal} {Rev. Mod. Phys.}\ }\textbf
  {\bibinfo {volume} {54}},\ \bibinfo {pages} {685} (\bibinfo {year}
  {1982})}\BibitemShut {NoStop}%
\bibitem [{\citenamefont {Guo}\ \emph {et~al.}(2020)\citenamefont {Guo},
  \citenamefont {Breum}, \citenamefont {Borregaard}, \citenamefont {Izumi},
  \citenamefont {Larsen}, \citenamefont {Gehring}, \citenamefont {Christandl},
  \citenamefont {Neergaard-Nielsen},\ and\ \citenamefont {Andersen}}]{Guo2020}%
  \BibitemOpen
  \bibfield  {author} {\bibinfo {author} {\bibfnamefont {X.}~\bibnamefont
  {Guo}}, \bibinfo {author} {\bibfnamefont {C.~R.}\ \bibnamefont {Breum}},
  \bibinfo {author} {\bibfnamefont {J.}~\bibnamefont {Borregaard}}, \bibinfo
  {author} {\bibfnamefont {S.}~\bibnamefont {Izumi}}, \bibinfo {author}
  {\bibfnamefont {M.~V.}\ \bibnamefont {Larsen}}, \bibinfo {author}
  {\bibfnamefont {T.}~\bibnamefont {Gehring}}, \bibinfo {author} {\bibfnamefont
  {M.}~\bibnamefont {Christandl}}, \bibinfo {author} {\bibfnamefont {J.~S.}\
  \bibnamefont {Neergaard-Nielsen}}, \ and\ \bibinfo {author} {\bibfnamefont
  {U.~L.}\ \bibnamefont {Andersen}},\ }\bibfield  {title} {\enquote {\bibinfo
  {title} {{Distributed quantum sensing in a continuous-variable entangled
  network}},}\ }\href {\doibase 10.1038/s41567-019-0743-x} {\bibfield
  {journal} {\bibinfo  {journal} {Nat. Phys.}\ }\textbf {\bibinfo {volume}
  {16}},\ \bibinfo {pages} {281} (\bibinfo {year} {2020})}\BibitemShut
  {NoStop}%
\bibitem [{\citenamefont {Picqu{\'{e}}}\ and\ \citenamefont
  {H{\"{a}}nsch}(2019)}]{combreview}%
  \BibitemOpen
  \bibfield  {author} {\bibinfo {author} {\bibfnamefont {N.}~\bibnamefont
  {Picqu{\'{e}}}}\ and\ \bibinfo {author} {\bibfnamefont {T.~W.}\ \bibnamefont
  {H{\"{a}}nsch}},\ }\bibfield  {title} {\enquote {\bibinfo {title} {{Frequency
  comb spectroscopy}},}\ }\href {\doibase 10.1038/s41566-018-0347-5} {\bibfield
   {journal} {\bibinfo  {journal} {Nat. Photon.}\ }\textbf {\bibinfo {volume}
  {13}},\ \bibinfo {pages} {146} (\bibinfo {year} {2019})}\BibitemShut
  {NoStop}%
\bibitem [{\citenamefont {Tran}\ \emph {et~al.}(2017)\citenamefont {Tran},
  \citenamefont {Sly},\ and\ \citenamefont {Conboy}}]{Tran2017}%
  \BibitemOpen
  \bibfield  {author} {\bibinfo {author} {\bibfnamefont {R.~J.}\ \bibnamefont
  {Tran}}, \bibinfo {author} {\bibfnamefont {K.~L.}\ \bibnamefont {Sly}}, \
  and\ \bibinfo {author} {\bibfnamefont {J.~C.}\ \bibnamefont {Conboy}},\
  }\bibfield  {title} {\enquote {\bibinfo {title} {{Applications of Surface
  Second Harmonic Generation in Biological Sensing}},}\ }\href {\doibase
  10.1146/annurev-anchem-071015-041453} {\bibfield  {journal} {\bibinfo
  {journal} {Annu. Rev. Anal. Chem}\ }\textbf {\bibinfo {volume} {10}},\
  \bibinfo {pages} {387} (\bibinfo {year} {2017})}\BibitemShut {NoStop}%
\bibitem [{\citenamefont {Kwiat}\ \emph
  {et~al.}(1995{\natexlab{a}})\citenamefont {Kwiat}, \citenamefont {Mattle},
  \citenamefont {Weinfurter}, \citenamefont {Zeilinger}, \citenamefont
  {Sergienko},\ and\ \citenamefont {Shih}}]{photonpair}%
  \BibitemOpen
  \bibfield  {author} {\bibinfo {author} {\bibfnamefont {P.~G.}\ \bibnamefont
  {Kwiat}}, \bibinfo {author} {\bibfnamefont {K.}~\bibnamefont {Mattle}},
  \bibinfo {author} {\bibfnamefont {H.}~\bibnamefont {Weinfurter}}, \bibinfo
  {author} {\bibfnamefont {A.}~\bibnamefont {Zeilinger}}, \bibinfo {author}
  {\bibfnamefont {A.~V.}\ \bibnamefont {Sergienko}}, \ and\ \bibinfo {author}
  {\bibfnamefont {Y.}~\bibnamefont {Shih}},\ }\bibfield  {title} {\enquote
  {\bibinfo {title} {New high-intensity source of polarization-entangled photon
  pairs},}\ }\href {\doibase 10.1103/PhysRevLett.75.4337} {\bibfield  {journal}
  {\bibinfo  {journal} {Phys. Rev. Lett.}\ }\textbf {\bibinfo {volume} {75}},\
  \bibinfo {pages} {4337} (\bibinfo {year} {1995}{\natexlab{a}})}\BibitemShut
  {NoStop}%
\bibitem [{\citenamefont {Kumar}(1990)}]{Kumar1990}%
  \BibitemOpen
  \bibfield  {author} {\bibinfo {author} {\bibfnamefont {P.}~\bibnamefont
  {Kumar}},\ }\bibfield  {title} {\enquote {\bibinfo {title} {{Quantum
  frequency conversion}},}\ }\href {\doibase 10.1364/ol.15.001476} {\bibfield
  {journal} {\bibinfo  {journal} {Opt. Lett.}\ }\textbf {\bibinfo {volume}
  {15}},\ \bibinfo {pages} {1476} (\bibinfo {year} {1990})}\BibitemShut
  {NoStop}%
\bibitem [{\citenamefont {Kok}\ and\ \citenamefont {Lovett}(2010)}]{Kok2010}%
  \BibitemOpen
  \bibfield  {author} {\bibinfo {author} {\bibfnamefont {P.}~\bibnamefont
  {Kok}}\ and\ \bibinfo {author} {\bibfnamefont {B.~W.}\ \bibnamefont
  {Lovett}},\ }\href {\doibase 10.1017/CBO9781139193658} {\emph {\bibinfo
  {title} {Introduction to Optical Quantum Information Processing}}}\ (\bibinfo
   {publisher} {Cambridge University Press},\ \bibinfo {address} {Cambridge},\
  \bibinfo {year} {2010})\ pp.\ \bibinfo {pages} {1--488}\BibitemShut {NoStop}%
\bibitem [{\citenamefont {Langford}\ \emph {et~al.}(2011)\citenamefont
  {Langford}, \citenamefont {Ramelow}, \citenamefont {Prevedel}, \citenamefont
  {Munro}, \citenamefont {Milburn},\ and\ \citenamefont
  {Zeilinger}}]{Langford2011}%
  \BibitemOpen
  \bibfield  {author} {\bibinfo {author} {\bibfnamefont {N.~K.}\ \bibnamefont
  {Langford}}, \bibinfo {author} {\bibfnamefont {S.}~\bibnamefont {Ramelow}},
  \bibinfo {author} {\bibfnamefont {R.}~\bibnamefont {Prevedel}}, \bibinfo
  {author} {\bibfnamefont {W.~J.}\ \bibnamefont {Munro}}, \bibinfo {author}
  {\bibfnamefont {G.~J.}\ \bibnamefont {Milburn}}, \ and\ \bibinfo {author}
  {\bibfnamefont {A.}~\bibnamefont {Zeilinger}},\ }\bibfield  {title} {\enquote
  {\bibinfo {title} {{Efficient quantum computing using coherent photon
  conversion}},}\ }\href {\doibase 10.1038/nature10463} {\bibfield  {journal}
  {\bibinfo  {journal} {Nature}\ }\textbf {\bibinfo {volume} {478}},\ \bibinfo
  {pages} {360} (\bibinfo {year} {2011})}\BibitemShut {NoStop}%
\bibitem [{\citenamefont {Ghimire}\ \emph {et~al.}(2011)\citenamefont
  {Ghimire}, \citenamefont {DiChiara}, \citenamefont {Sistrunk}, \citenamefont
  {Agostini}, \citenamefont {DiMauro},\ and\ \citenamefont
  {Reis}}]{Ghimire2011}%
  \BibitemOpen
  \bibfield  {author} {\bibinfo {author} {\bibfnamefont {S.}~\bibnamefont
  {Ghimire}}, \bibinfo {author} {\bibfnamefont {A.~D.}\ \bibnamefont
  {DiChiara}}, \bibinfo {author} {\bibfnamefont {E.}~\bibnamefont {Sistrunk}},
  \bibinfo {author} {\bibfnamefont {P.}~\bibnamefont {Agostini}}, \bibinfo
  {author} {\bibfnamefont {L.~F.}\ \bibnamefont {DiMauro}}, \ and\ \bibinfo
  {author} {\bibfnamefont {D.~A.}\ \bibnamefont {Reis}},\ }\bibfield  {title}
  {\enquote {\bibinfo {title} {{Observation of high-order harmonic generation
  in a bulk crystal}},}\ }\href {\doibase 10.1038/nphys1847} {\bibfield
  {journal} {\bibinfo  {journal} {Nat. Phys.}\ }\textbf {\bibinfo {volume}
  {7}},\ \bibinfo {pages} {138} (\bibinfo {year} {2011})}\BibitemShut {NoStop}%
\bibitem [{\citenamefont {Cox}\ \emph {et~al.}(2017)\citenamefont {Cox},
  \citenamefont {Marini},\ and\ \citenamefont {de~Abajo}}]{Cox2017}%
  \BibitemOpen
  \bibfield  {author} {\bibinfo {author} {\bibfnamefont {J.~D.}\ \bibnamefont
  {Cox}}, \bibinfo {author} {\bibfnamefont {A.}~\bibnamefont {Marini}}, \ and\
  \bibinfo {author} {\bibfnamefont {F.~J.~G.}\ \bibnamefont {de~Abajo}},\
  }\bibfield  {title} {\enquote {\bibinfo {title} {{Plasmon-assisted
  high-harmonic generation in graphene}},}\ }\href {\doibase
  10.1038/ncomms14380} {\bibfield  {journal} {\bibinfo  {journal} {Nat.
  Commun.}\ }\textbf {\bibinfo {volume} {8}},\ \bibinfo {pages} {14380}
  (\bibinfo {year} {2017})}\BibitemShut {NoStop}%
\bibitem [{\citenamefont {Strekalov}\ \emph {et~al.}(2016)\citenamefont
  {Strekalov}, \citenamefont {Marquardt}, \citenamefont {Matsko}, \citenamefont
  {Schwefel},\ and\ \citenamefont {Leuchs}}]{Strekalov2016}%
  \BibitemOpen
  \bibfield  {author} {\bibinfo {author} {\bibfnamefont {D.~V.}\ \bibnamefont
  {Strekalov}}, \bibinfo {author} {\bibfnamefont {C.}~\bibnamefont
  {Marquardt}}, \bibinfo {author} {\bibfnamefont {A.~B.}\ \bibnamefont
  {Matsko}}, \bibinfo {author} {\bibfnamefont {H.~G.~L.}\ \bibnamefont
  {Schwefel}}, \ and\ \bibinfo {author} {\bibfnamefont {G.}~\bibnamefont
  {Leuchs}},\ }\bibfield  {title} {\enquote {\bibinfo {title} {{Nonlinear and
  quantum optics with whispering gallery resonators}},}\ }\href {\doibase
  10.1088/2040-8978/18/12/123002} {\bibfield  {journal} {\bibinfo  {journal}
  {J. Opt.}\ }\textbf {\bibinfo {volume} {18}},\ \bibinfo {pages} {123002}
  (\bibinfo {year} {2016})}\BibitemShut {NoStop}%
\bibitem [{\citenamefont {Breunig}(2016)}]{Breunig2016}%
  \BibitemOpen
  \bibfield  {author} {\bibinfo {author} {\bibfnamefont {I.}~\bibnamefont
  {Breunig}},\ }\bibfield  {title} {\enquote {\bibinfo {title} {{Three-wave
  mixing in whispering gallery resonators}},}\ }\href {\doibase
  10.1002/lpor.201600038} {\bibfield  {journal} {\bibinfo  {journal} {Laser
  Photon. Rev.}\ }\textbf {\bibinfo {volume} {10}},\ \bibinfo {pages} {569}
  (\bibinfo {year} {2016})}\BibitemShut {NoStop}%
\bibitem [{\citenamefont {Elshaari}\ \emph {et~al.}(2020)\citenamefont
  {Elshaari}, \citenamefont {Pernice}, \citenamefont {Srinivasan},
  \citenamefont {Benson},\ and\ \citenamefont {Zwiller}}]{Elshaari2020}%
  \BibitemOpen
  \bibfield  {author} {\bibinfo {author} {\bibfnamefont {A.~W.}\ \bibnamefont
  {Elshaari}}, \bibinfo {author} {\bibfnamefont {W.}~\bibnamefont {Pernice}},
  \bibinfo {author} {\bibfnamefont {K.}~\bibnamefont {Srinivasan}}, \bibinfo
  {author} {\bibfnamefont {O.}~\bibnamefont {Benson}}, \ and\ \bibinfo {author}
  {\bibfnamefont {V.}~\bibnamefont {Zwiller}},\ }\bibfield  {title} {\enquote
  {\bibinfo {title} {{Hybrid integrated quantum photonic circuits}},}\ }\href
  {\doibase 10.1038/s41566-020-0609-x} {\bibfield  {journal} {\bibinfo
  {journal} {Nat. Photon.}\ }\textbf {\bibinfo {volume} {14}},\ \bibinfo
  {pages} {285} (\bibinfo {year} {2020})}\BibitemShut {NoStop}%
\bibitem [{\citenamefont {Xiao}\ \emph {et~al.}(2020)\citenamefont {Xiao},
  \citenamefont {Zou}, \citenamefont {Gong},\ and\ \citenamefont
  {Yang}}]{Xiao2020}%
  \BibitemOpen
  \bibfield  {author} {\bibinfo {author} {\bibfnamefont {Y.~F.}\ \bibnamefont
  {Xiao}}, \bibinfo {author} {\bibfnamefont {C.~L.}\ \bibnamefont {Zou}},
  \bibinfo {author} {\bibfnamefont {Q.}~\bibnamefont {Gong}}, \ and\ \bibinfo
  {author} {\bibfnamefont {L.}~\bibnamefont {Yang}},\ }\href {\doibase
  10.1142/8964} {\emph {\bibinfo {title} {Ultra-high-Q Optical
  Microcavities}}}\ (\bibinfo  {publisher} {World Scientific},\ \bibinfo {year}
  {2020})\ pp.\ \bibinfo {pages} {1--403}\BibitemShut {NoStop}%
\bibitem [{\citenamefont {Lu}\ \emph {et~al.}(2021)\citenamefont {Lu},
  \citenamefont {{Al Sayem}}, \citenamefont {Gong}, \citenamefont {Surya},
  \citenamefont {Zou},\ and\ \citenamefont {Tang}}]{Lu2021}%
  \BibitemOpen
  \bibfield  {author} {\bibinfo {author} {\bibfnamefont {J.}~\bibnamefont
  {Lu}}, \bibinfo {author} {\bibfnamefont {A.}~\bibnamefont {{Al Sayem}}},
  \bibinfo {author} {\bibfnamefont {Z.}~\bibnamefont {Gong}}, \bibinfo {author}
  {\bibfnamefont {J.~B.}\ \bibnamefont {Surya}}, \bibinfo {author}
  {\bibfnamefont {C.-L.}\ \bibnamefont {Zou}}, \ and\ \bibinfo {author}
  {\bibfnamefont {H.~X.}\ \bibnamefont {Tang}},\ }\bibfield  {title} {\enquote
  {\bibinfo {title} {{Ultralow-threshold thin-film lithium niobate optical
  parametric oscillator}},}\ }\href {\doibase 10.1364/OPTICA.418984} {\bibfield
   {journal} {\bibinfo  {journal} {Optica}\ }\textbf {\bibinfo {volume} {8}},\
  \bibinfo {pages} {539} (\bibinfo {year} {2021})}\BibitemShut {NoStop}%
\bibitem [{\citenamefont {Marty}\ \emph {et~al.}(2021)\citenamefont {Marty},
  \citenamefont {Combri{\'{e}}}, \citenamefont {Raineri},\ and\ \citenamefont
  {{De Rossi}}}]{Marty2021}%
  \BibitemOpen
  \bibfield  {author} {\bibinfo {author} {\bibfnamefont {G.}~\bibnamefont
  {Marty}}, \bibinfo {author} {\bibfnamefont {S.}~\bibnamefont
  {Combri{\'{e}}}}, \bibinfo {author} {\bibfnamefont {F.}~\bibnamefont
  {Raineri}}, \ and\ \bibinfo {author} {\bibfnamefont {A.}~\bibnamefont {{De
  Rossi}}},\ }\bibfield  {title} {\enquote {\bibinfo {title} {{Photonic crystal
  optical parametric oscillator}},}\ }\href {\doibase
  10.1038/s41566-020-00737-z} {\bibfield  {journal} {\bibinfo  {journal} {Nat.
  Photon.}\ }\textbf {\bibinfo {volume} {15}},\ \bibinfo {pages} {53} (\bibinfo
  {year} {2021})}\BibitemShut {NoStop}%
\bibitem [{\citenamefont {Guo}\ \emph {et~al.}(2016)\citenamefont {Guo},
  \citenamefont {Zou},\ and\ \citenamefont {Tang}}]{Guo2016}%
  \BibitemOpen
  \bibfield  {author} {\bibinfo {author} {\bibfnamefont {X.}~\bibnamefont
  {Guo}}, \bibinfo {author} {\bibfnamefont {C.}~\bibnamefont {Zou}}, \ and\
  \bibinfo {author} {\bibfnamefont {H.}~\bibnamefont {Tang}},\ }\bibfield
  {title} {\enquote {\bibinfo {title} {{Second-harmonic generation in aluminum
  nitride microrings with 2500 {\%}/ W conversion efficiency}},}\ }\href
  {\doibase 10.1364/OPTICA.3.001126} {\bibfield  {journal} {\bibinfo  {journal}
  {Optica}\ }\textbf {\bibinfo {volume} {3}},\ \bibinfo {pages} {1126}
  (\bibinfo {year} {2016})}\BibitemShut {NoStop}%
\bibitem [{\citenamefont {Bruch}\ \emph {et~al.}(2021)\citenamefont {Bruch},
  \citenamefont {Liu}, \citenamefont {Gong}, \citenamefont {Surya},
  \citenamefont {Li}, \citenamefont {Zou},\ and\ \citenamefont
  {Tang}}]{Bruch2021}%
  \BibitemOpen
  \bibfield  {author} {\bibinfo {author} {\bibfnamefont {A.~W.}\ \bibnamefont
  {Bruch}}, \bibinfo {author} {\bibfnamefont {X.}~\bibnamefont {Liu}}, \bibinfo
  {author} {\bibfnamefont {Z.}~\bibnamefont {Gong}}, \bibinfo {author}
  {\bibfnamefont {J.~B.}\ \bibnamefont {Surya}}, \bibinfo {author}
  {\bibfnamefont {M.}~\bibnamefont {Li}}, \bibinfo {author} {\bibfnamefont
  {C.-L.}\ \bibnamefont {Zou}}, \ and\ \bibinfo {author} {\bibfnamefont
  {H.~X.}\ \bibnamefont {Tang}},\ }\bibfield  {title} {\enquote {\bibinfo
  {title} {{Pockels soliton microcomb}},}\ }\href {\doibase
  10.1038/s41566-020-00704-8} {\bibfield  {journal} {\bibinfo  {journal} {Nat.
  Photon.}\ }\textbf {\bibinfo {volume} {15}},\ \bibinfo {pages} {21} (\bibinfo
  {year} {2021})}\BibitemShut {NoStop}%
\bibitem [{\citenamefont {Kwiat}\ \emph
  {et~al.}(1995{\natexlab{b}})\citenamefont {Kwiat}, \citenamefont {Mattle},
  \citenamefont {Weinfurter}, \citenamefont {Zeilinger}, \citenamefont
  {Sergienko},\ and\ \citenamefont {Shih}}]{Kwiat1995}%
  \BibitemOpen
  \bibfield  {author} {\bibinfo {author} {\bibfnamefont {P.~G.}\ \bibnamefont
  {Kwiat}}, \bibinfo {author} {\bibfnamefont {K.}~\bibnamefont {Mattle}},
  \bibinfo {author} {\bibfnamefont {H.}~\bibnamefont {Weinfurter}}, \bibinfo
  {author} {\bibfnamefont {A.}~\bibnamefont {Zeilinger}}, \bibinfo {author}
  {\bibfnamefont {A.~V.}\ \bibnamefont {Sergienko}}, \ and\ \bibinfo {author}
  {\bibfnamefont {Y.}~\bibnamefont {Shih}},\ }\bibfield  {title} {\enquote
  {\bibinfo {title} {{New High-Intensity Source of Polarization-Entangled
  Photon Pairs}},}\ }\href {\doibase 10.1103/PhysRevLett.75.4337} {\bibfield
  {journal} {\bibinfo  {journal} {Phys. Rev. Lett.}\ }\textbf {\bibinfo
  {volume} {75}},\ \bibinfo {pages} {4337} (\bibinfo {year}
  {1995}{\natexlab{b}})}\BibitemShut {NoStop}%
\bibitem [{\citenamefont {Ma}\ \emph {et~al.}(2012)\citenamefont {Ma},
  \citenamefont {Herbst}, \citenamefont {Scheidl}, \citenamefont {Wang},
  \citenamefont {Kropatschek}, \citenamefont {Naylor}, \citenamefont
  {Wittmann}, \citenamefont {Mech}, \citenamefont {Kofler}, \citenamefont
  {Anisimova}, \citenamefont {Makarov}, \citenamefont {Jennewein},
  \citenamefont {Ursin},\ and\ \citenamefont {Zeilinger}}]{Ma2012}%
  \BibitemOpen
  \bibfield  {author} {\bibinfo {author} {\bibfnamefont {X.~S.}\ \bibnamefont
  {Ma}}, \bibinfo {author} {\bibfnamefont {T.}~\bibnamefont {Herbst}}, \bibinfo
  {author} {\bibfnamefont {T.}~\bibnamefont {Scheidl}}, \bibinfo {author}
  {\bibfnamefont {D.}~\bibnamefont {Wang}}, \bibinfo {author} {\bibfnamefont
  {S.}~\bibnamefont {Kropatschek}}, \bibinfo {author} {\bibfnamefont
  {W.}~\bibnamefont {Naylor}}, \bibinfo {author} {\bibfnamefont
  {B.}~\bibnamefont {Wittmann}}, \bibinfo {author} {\bibfnamefont
  {A.}~\bibnamefont {Mech}}, \bibinfo {author} {\bibfnamefont {J.}~\bibnamefont
  {Kofler}}, \bibinfo {author} {\bibfnamefont {E.}~\bibnamefont {Anisimova}},
  \bibinfo {author} {\bibfnamefont {V.}~\bibnamefont {Makarov}}, \bibinfo
  {author} {\bibfnamefont {T.}~\bibnamefont {Jennewein}}, \bibinfo {author}
  {\bibfnamefont {R.}~\bibnamefont {Ursin}}, \ and\ \bibinfo {author}
  {\bibfnamefont {A.}~\bibnamefont {Zeilinger}},\ }\bibfield  {title} {\enquote
  {\bibinfo {title} {{Quantum teleportation over 143 kilometres using active
  feed-forward}},}\ }\href {\doibase 10.1038/nature11472} {\bibfield  {journal}
  {\bibinfo  {journal} {Nature}\ }\textbf {\bibinfo {volume} {489}},\ \bibinfo
  {pages} {269} (\bibinfo {year} {2012})}\BibitemShut {NoStop}%
\bibitem [{\citenamefont {Lu}\ \emph {et~al.}(2019)\citenamefont {Lu},
  \citenamefont {Li}, \citenamefont {Westly}, \citenamefont {Moille},
  \citenamefont {Singh}, \citenamefont {Anant},\ and\ \citenamefont
  {Srinivasan}}]{Kartik2019}%
  \BibitemOpen
  \bibfield  {author} {\bibinfo {author} {\bibfnamefont {X.}~\bibnamefont
  {Lu}}, \bibinfo {author} {\bibfnamefont {Q.}~\bibnamefont {Li}}, \bibinfo
  {author} {\bibfnamefont {D.~A.}\ \bibnamefont {Westly}}, \bibinfo {author}
  {\bibfnamefont {G.}~\bibnamefont {Moille}}, \bibinfo {author} {\bibfnamefont
  {A.}~\bibnamefont {Singh}}, \bibinfo {author} {\bibfnamefont
  {V.}~\bibnamefont {Anant}}, \ and\ \bibinfo {author} {\bibfnamefont
  {K.}~\bibnamefont {Srinivasan}},\ }\bibfield  {title} {\enquote {\bibinfo
  {title} {Chip-integrated visible--telecom entangled photon pair source for
  quantum communication},}\ }\href
  {http://www.nature.com/articles/s41567-018-0394-3} {\bibfield  {journal}
  {\bibinfo  {journal} {Nat. Phys.}\ }\textbf {\bibinfo {volume} {15}},\
  \bibinfo {pages} {373} (\bibinfo {year} {2019})}\BibitemShut {NoStop}%
\bibitem [{\citenamefont {Franson}(1989)}]{Franson1989}%
  \BibitemOpen
  \bibfield  {author} {\bibinfo {author} {\bibfnamefont {J.~D.}\ \bibnamefont
  {Franson}},\ }\bibfield  {title} {\enquote {\bibinfo {title} {{Bell
  inequality for position and time}},}\ }\href {\doibase
  10.1103/PhysRevLett.62.2205} {\bibfield  {journal} {\bibinfo  {journal}
  {Phys. Rev. Lett.}\ }\textbf {\bibinfo {volume} {62}},\ \bibinfo {pages}
  {2205} (\bibinfo {year} {1989})}\BibitemShut {NoStop}%
\bibitem [{\citenamefont {Yu}\ \emph {et~al.}(2020)\citenamefont {Yu},
  \citenamefont {Ma}, \citenamefont {Luo}, \citenamefont {Jing}, \citenamefont
  {Sun}, \citenamefont {Fang}, \citenamefont {Yang}, \citenamefont {Liu},
  \citenamefont {Zheng}, \citenamefont {Xie}, \citenamefont {Zhang},
  \citenamefont {You}, \citenamefont {Wang}, \citenamefont {Chen},
  \citenamefont {Zhang}, \citenamefont {Bao},\ and\ \citenamefont
  {Pan}}]{Yu2020}%
  \BibitemOpen
  \bibfield  {author} {\bibinfo {author} {\bibfnamefont {Y.}~\bibnamefont
  {Yu}}, \bibinfo {author} {\bibfnamefont {F.}~\bibnamefont {Ma}}, \bibinfo
  {author} {\bibfnamefont {X.-Y.}\ \bibnamefont {Luo}}, \bibinfo {author}
  {\bibfnamefont {B.}~\bibnamefont {Jing}}, \bibinfo {author} {\bibfnamefont
  {P.-F.}\ \bibnamefont {Sun}}, \bibinfo {author} {\bibfnamefont {R.-Z.}\
  \bibnamefont {Fang}}, \bibinfo {author} {\bibfnamefont {C.-W.}\ \bibnamefont
  {Yang}}, \bibinfo {author} {\bibfnamefont {H.}~\bibnamefont {Liu}}, \bibinfo
  {author} {\bibfnamefont {M.-Y.}\ \bibnamefont {Zheng}}, \bibinfo {author}
  {\bibfnamefont {X.-P.}\ \bibnamefont {Xie}}, \bibinfo {author} {\bibfnamefont
  {W.-J.}\ \bibnamefont {Zhang}}, \bibinfo {author} {\bibfnamefont {L.-X.}\
  \bibnamefont {You}}, \bibinfo {author} {\bibfnamefont {Z.}~\bibnamefont
  {Wang}}, \bibinfo {author} {\bibfnamefont {T.-Y.}\ \bibnamefont {Chen}},
  \bibinfo {author} {\bibfnamefont {Q.}~\bibnamefont {Zhang}}, \bibinfo
  {author} {\bibfnamefont {X.-H.}\ \bibnamefont {Bao}}, \ and\ \bibinfo
  {author} {\bibfnamefont {J.-W.}\ \bibnamefont {Pan}},\ }\bibfield  {title}
  {\enquote {\bibinfo {title} {{Entanglement of two quantum memories via fibres
  over dozens of kilometres}},}\ }\href {\doibase 10.1038/s41586-020-1976-7}
  {\bibfield  {journal} {\bibinfo  {journal} {Nature}\ }\textbf {\bibinfo
  {volume} {578}},\ \bibinfo {pages} {240} (\bibinfo {year}
  {2020})}\BibitemShut {NoStop}%
\bibitem [{\citenamefont {Shan}\ \emph {et~al.}(2021)\citenamefont {Shan},
  \citenamefont {Ye}, \citenamefont {Chu}, \citenamefont {Lee}, \citenamefont
  {Park}, \citenamefont {Balents},\ and\ \citenamefont
  {Hsieh}}]{shan2021giant}%
  \BibitemOpen
  \bibfield  {author} {\bibinfo {author} {\bibfnamefont {J.-Y.}\ \bibnamefont
  {Shan}}, \bibinfo {author} {\bibfnamefont {M.}~\bibnamefont {Ye}}, \bibinfo
  {author} {\bibfnamefont {H.}~\bibnamefont {Chu}}, \bibinfo {author}
  {\bibfnamefont {S.}~\bibnamefont {Lee}}, \bibinfo {author} {\bibfnamefont
  {J.-G.}\ \bibnamefont {Park}}, \bibinfo {author} {\bibfnamefont
  {L.}~\bibnamefont {Balents}}, \ and\ \bibinfo {author} {\bibfnamefont
  {D.}~\bibnamefont {Hsieh}},\ }\bibfield  {title} {\enquote {\bibinfo {title}
  {Giant modulation of optical nonlinearity by floquet engineering},}\ }\href
  {\doibase https://doi.org/10.1038/s41586-021-04051-8} {\bibfield  {journal}
  {\bibinfo  {journal} {Nature}\ }\textbf {\bibinfo {volume} {600}},\ \bibinfo
  {pages} {235} (\bibinfo {year} {2021})}\BibitemShut {NoStop}%
\bibitem [{\citenamefont {Llewellyn}\ \emph {et~al.}(2020)\citenamefont
  {Llewellyn}, \citenamefont {Ding}, \citenamefont {Faruque}, \citenamefont
  {Paesani}, \citenamefont {Bacco}, \citenamefont {Santagati}, \citenamefont
  {Qian}, \citenamefont {Li}, \citenamefont {Xiao}, \citenamefont {Huber},
  \citenamefont {Malik}, \citenamefont {Sinclair}, \citenamefont {Zhou},
  \citenamefont {Rottwitt}, \citenamefont {O'Brien}, \citenamefont {Rarity},
  \citenamefont {Gong}, \citenamefont {Oxenlowe}, \citenamefont {Wang},\ and\
  \citenamefont {Thompson}}]{Llewellyn2020}%
  \BibitemOpen
  \bibfield  {author} {\bibinfo {author} {\bibfnamefont {D.}~\bibnamefont
  {Llewellyn}}, \bibinfo {author} {\bibfnamefont {Y.}~\bibnamefont {Ding}},
  \bibinfo {author} {\bibfnamefont {I.~I.}\ \bibnamefont {Faruque}}, \bibinfo
  {author} {\bibfnamefont {S.}~\bibnamefont {Paesani}}, \bibinfo {author}
  {\bibfnamefont {D.}~\bibnamefont {Bacco}}, \bibinfo {author} {\bibfnamefont
  {R.}~\bibnamefont {Santagati}}, \bibinfo {author} {\bibfnamefont {Y.-J.}\
  \bibnamefont {Qian}}, \bibinfo {author} {\bibfnamefont {Y.}~\bibnamefont
  {Li}}, \bibinfo {author} {\bibfnamefont {Y.-F.}\ \bibnamefont {Xiao}},
  \bibinfo {author} {\bibfnamefont {M.}~\bibnamefont {Huber}}, \bibinfo
  {author} {\bibfnamefont {M.}~\bibnamefont {Malik}}, \bibinfo {author}
  {\bibfnamefont {G.~F.}\ \bibnamefont {Sinclair}}, \bibinfo {author}
  {\bibfnamefont {X.}~\bibnamefont {Zhou}}, \bibinfo {author} {\bibfnamefont
  {K.}~\bibnamefont {Rottwitt}}, \bibinfo {author} {\bibfnamefont {J.~L.}\
  \bibnamefont {O'Brien}}, \bibinfo {author} {\bibfnamefont {J.~G.}\
  \bibnamefont {Rarity}}, \bibinfo {author} {\bibfnamefont {Q.}~\bibnamefont
  {Gong}}, \bibinfo {author} {\bibfnamefont {L.~K.}\ \bibnamefont {Oxenlowe}},
  \bibinfo {author} {\bibfnamefont {J.}~\bibnamefont {Wang}}, \ and\ \bibinfo
  {author} {\bibfnamefont {M.~G.}\ \bibnamefont {Thompson}},\ }\bibfield
  {title} {\enquote {\bibinfo {title} {{Chip-to-chip quantum teleportation and
  multi-photon entanglement in silicon}},}\ }\href {\doibase
  10.1038/s41567-019-0727-x} {\bibfield  {journal} {\bibinfo  {journal} {Nat.
  Phys.}\ }\textbf {\bibinfo {volume} {16}},\ \bibinfo {pages} {148} (\bibinfo
  {year} {2020})}\BibitemShut {NoStop}%
\bibitem [{\citenamefont {Kuo}\ \emph {et~al.}(2014)\citenamefont {Kuo},
  \citenamefont {Bravo-Abad},\ and\ \citenamefont {Solomon}}]{Kuo2014}%
  \BibitemOpen
  \bibfield  {author} {\bibinfo {author} {\bibfnamefont {P.~S.}\ \bibnamefont
  {Kuo}}, \bibinfo {author} {\bibfnamefont {J.}~\bibnamefont {Bravo-Abad}}, \
  and\ \bibinfo {author} {\bibfnamefont {G.~S.}\ \bibnamefont {Solomon}},\
  }\bibfield  {title} {\enquote {\bibinfo {title} {{Second-harmonic generation
  using -quasi-phasematching in a GaAs whispering-gallery-mode microcavity}},}\
  }\href {\doibase 10.1038/ncomms4109} {\bibfield  {journal} {\bibinfo
  {journal} {Nat. Commun.}\ }\textbf {\bibinfo {volume} {5}},\ \bibinfo {pages}
  {3109} (\bibinfo {year} {2014})}\BibitemShut {NoStop}%
\bibitem [{\citenamefont {Chen}\ \emph {et~al.}(2019)\citenamefont {Chen},
  \citenamefont {Ma}, \citenamefont {Sua}, \citenamefont {Li}, \citenamefont
  {Tang},\ and\ \citenamefont {Huang}}]{ChenLN2019}%
  \BibitemOpen
  \bibfield  {author} {\bibinfo {author} {\bibfnamefont {J.-Y.}\ \bibnamefont
  {Chen}}, \bibinfo {author} {\bibfnamefont {Z.-H.}\ \bibnamefont {Ma}},
  \bibinfo {author} {\bibfnamefont {Y.~M.}\ \bibnamefont {Sua}}, \bibinfo
  {author} {\bibfnamefont {Z.}~\bibnamefont {Li}}, \bibinfo {author}
  {\bibfnamefont {C.}~\bibnamefont {Tang}}, \ and\ \bibinfo {author}
  {\bibfnamefont {Y.-P.}\ \bibnamefont {Huang}},\ }\bibfield  {title} {\enquote
  {\bibinfo {title} {Ultra-efficient frequency conversion in
  quasi-phase-matched lithium niobate microrings},}\ }\href
  {http://www.osapublishing.org/optica/abstract.cfm?URI=optica-6-9-1244}
  {\bibfield  {journal} {\bibinfo  {journal} {Optica}\ }\textbf {\bibinfo
  {volume} {6}},\ \bibinfo {pages} {1244} (\bibinfo {year} {2019})}\BibitemShut
  {NoStop}%
\bibitem [{\citenamefont {Lu}\ \emph {et~al.}(2020)\citenamefont {Lu},
  \citenamefont {Li}, \citenamefont {Zou}, \citenamefont {{Al Sayem}},\ and\
  \citenamefont {Tang}}]{Lu2020}%
  \BibitemOpen
  \bibfield  {author} {\bibinfo {author} {\bibfnamefont {J.}~\bibnamefont
  {Lu}}, \bibinfo {author} {\bibfnamefont {M.}~\bibnamefont {Li}}, \bibinfo
  {author} {\bibfnamefont {C.-L.}\ \bibnamefont {Zou}}, \bibinfo {author}
  {\bibfnamefont {A.}~\bibnamefont {{Al Sayem}}}, \ and\ \bibinfo {author}
  {\bibfnamefont {H.~X.}\ \bibnamefont {Tang}},\ }\bibfield  {title} {\enquote
  {\bibinfo {title} {{Toward 1\% single-photon anharmonicity with periodically
  poled lithium niobate microring resonators}},}\ }\href {\doibase
  10.1364/optica.403931} {\bibfield  {journal} {\bibinfo  {journal} {Optica}\
  }\textbf {\bibinfo {volume} {7}},\ \bibinfo {pages} {1654} (\bibinfo {year}
  {2020})}\BibitemShut {NoStop}%
\end{thebibliography}
\end{document}